\documentclass[submission,Phys]{SciPost}

\hypersetup{
    colorlinks,
    linkcolor={red!50!black},
    citecolor={blue!50!black},
    urlcolor={blue!80!black}
}

\usepackage[bitstream-charter]{mathdesign}
\usepackage{graphicx} 
\usepackage{amsmath}
\newcommand{\be}{\begin{equation}}
\newcommand{\ee}{\end{equation}}
\newcommand{\bea}{\begin{eqnarray}}
\newcommand{\eea}{\end{eqnarray}}
\newcommand{\up}{\uparrow}

\def\sfix#1{\texorpdfstring{#1}{Lg}}
\def\doi{http://dx.doi.org/}
\def\nn{\nonumber\\}

\def\fr#1{(\ref{#1})}

\DeclareSymbolFont{usualmathcal}{OMS}{cmsy}{m}{n}
\DeclareSymbolFontAlphabet{\mathcal}{usualmathcal}

\fancypagestyle{SPstyle}{
\fancyhf{}
\lhead{\colorbox{scipostblue}{\bf \color{white} ~SciPost Physics }}
\rhead{{\bf \color{scipostdeepblue} ~Submission }}

\fancyfoot[C]{\textbf{\thepage}}
}

\begin{document}

\pagestyle{SPstyle}

\begin{center}{\Large \textbf{\color{scipostdeepblue}{
Criticality enabled long-range order in a U(1)-symmetric spin-1 Heisenberg chain with biquadratic interactions\\
}}}\end{center}

\begin{center}\textbf{
Natalia Chepiga\textsuperscript{$\star$} and
Fabian H.L. Essler\textsuperscript{$\dagger$}}
\end{center}

\begin{center}
{\bf 1} 
The Rudolf Peierls Centre for Theoretical Physics, Oxford
  University, Oxford OX1 3PU, United Kingdom
\\[\baselineskip]
$\star$ \href{mailto:email1}{\small natalia.chepiga@physics.ox.ac.uk}\,,\quad
$\dagger$ \href{mailto:email2}{\small fab@thphys.ox.ac.uk }
\end{center}

\section*{\color{scipostdeepblue}{Abstract}}
\textbf{\boldmath{
In a recent paper Nahum argued on the basis of a renormalization-group analysis that, contrary to standard lore, ground states of 1D spin chains with short-range interactions can spontaneously break U(1) ``easy-plane'' spin rotation symmetry. True long-range order of $(S^x,S^y)$ arises at the phase transition between two quasi-long-range-ordered phases. Here we present detailed numerical results for an anisotropic spin-1 chain model.
 We find a magnetization exponent $\beta\approx0.29$, consistent with the $\epsilon$-expansion prediction $\beta\approx0.28$, a divergence of the Luttinger parameter on approaching the transition from both sides with an exponent that we relate to Nahum's transition, and finite-size spectra consistent with a dynamical critical exponent $z\approx2$. Our results support the critical behavior predicted by Nahum's field theory.}
}

\vspace{\baselineskip}

\vspace{10pt}
\noindent\rule{\textwidth}{1pt}
\tableofcontents
\noindent\rule{\textwidth}{1pt}
\vspace{10pt}

\section{Introduction}
\label{sec:intro}
\begin{figure}[ht]
    \centering
    \includegraphics[width=0.7\columnwidth]{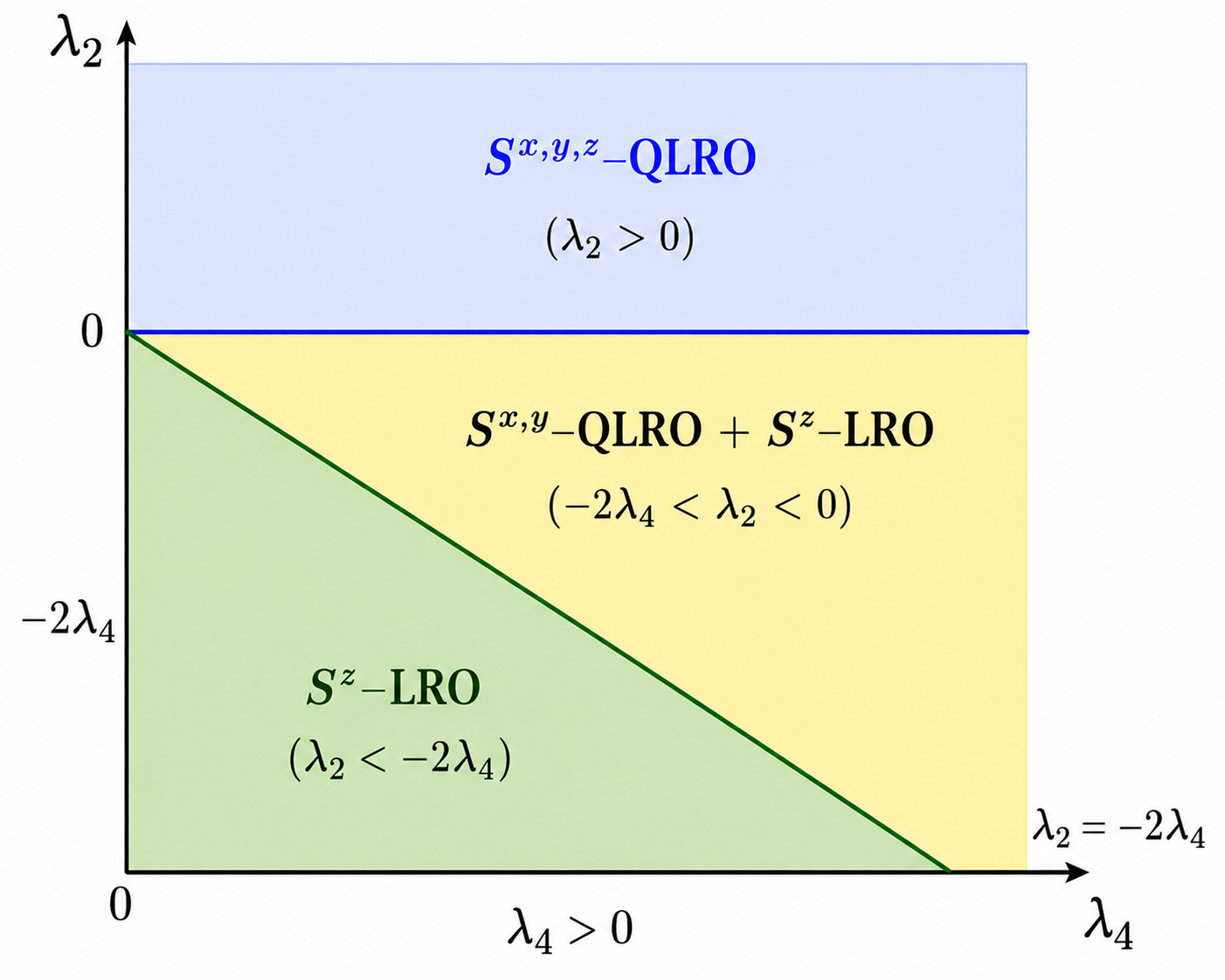}
    \caption{Phase diagram of the field theory \fr{SFT} according to Ref.~\cite{nahum2025continuous}.}
    \label{fig:PDFT}
\end{figure}

According to a widely held belief rooted in the seminal work of Mermin, Wagner, Hohenberg and others \cite{mermin1966absence,hohenberg1967existence,mermin1967absence,
garrison1972absence,mcbryan1977decay,klein1981absence,frohlich1981absence,shastry1992bounds,halperin2019hohenberg}, continuous symmetries cannot be spontaneously broken in the ground states of one dimensional quantum spin chains with local interactions, with the exception of cases in which the order parameter commutes with the Hamiltonian. The latter qualification accounts e.g. for the case of the spin-1/2 Heisenberg ferromagnet
\be
{\mathcal{H}=-J\sum_i\boldsymbol{S}_i\cdot\boldsymbol{S}_{i+1}}\ ,
\ee
which on a finite chain of $N$ sites has an $N+1$-fold degenerate ground state
\be
(S^-)^n|\up\dots\up\rangle\ ,\quad n=0,\dots N.
\ee
Here $S^-$ denotes the global SU(2) spin lowering operator. It is straightforward to see that applying an infinitesimal symmetry-breaking magnetic field along the $z$-direction and taking the order of limits appropriate for probing spontaneous symmetry breaking gives a non-zero magnetization per site
\be
m=\lim_{\epsilon\to 0}\lim_{N\to\infty}\lim_{T\to 0}\frac{{\rm Tr}\big(e^{-{\mathcal{H}/T}}S^z\big)}{{\rm Tr}\big(e^{-{\mathcal{H}/T}}\big)}=\frac{1}{2}\ .
\ee
The Heisenberg ferromagnet is peculiar because the order parameter commutes with the Hamiltonian $[\mathcal{H},S^z]=0$. The analogous situation with superconducting order associated with an eta-pairing like SU(2) symmetry exists in extended Hubbard models \cite{essler1992new}.  
Several recent studies established that spontaneous symmetry breaking of continuous symmetries is possible even when the order parameter does not commute with the Hamiltonian \cite{watanabe2024critical,fava2024heisenberg}, but required either a fine-tuning of parameters and associated large ground-state degeneracy or non-abelian symmetry. Ref.~\cite{nahum2025continuous} considered a generic quantum spin chain with a continuous abelian U(1) symmetry and showed by means of renormalization-group methods that true long-range order occurs at a zero-temperature phase transition that can be reached by tuning one parameter. The starting point of the RG analysis is the imaginary-time action of a perturbed SO(3) ferromagnet 
\begin{align}
S&=\int_0^\beta d\tau\int dx \Big(-iS(1-\psi)\partial_t\theta-\frac{JS^2}{2}(\partial_x\boldsymbol{n})^2+V(\psi)\Big)\ ,\nn
\boldsymbol{n}&=(\sqrt{1-\psi^2}\cos\theta,\sqrt{1-\psi^2}\sin\theta,\psi)\ ,\quad
V(\psi)=\lambda_2\psi^2+\lambda_4\psi^4\ .
\label{SFT}
\end{align}
In the vicinity of the phase transition of interest this gives rise to the following Lagrangian density 
\be
\mathcal{L}=i\psi\partial_\tau\theta+\frac{1}{2}(\partial_x\theta)^2+\frac{1}{2}(\partial_x\psi)^2+V(\psi)\ .
\label{Lagr}
\ee
The mean-field ground-state phase diagram resulting from \fr{Lagr} proposed in Ref.~\cite{nahum2025continuous} is shown in Fig. \ref{fig:PDFT}
There are three distinct phases:
\begin{enumerate}
\item{} An unmagnetized Luttinger-liquid phase, which in the mean-field approximation occurs at $\lambda_2>0$. Here the spin-spin correlation functions along all three directions display power-law decays. The dynamical critical exponent is $z=1$.
\item{} A partially magnetized Luttinger-liquid phase. Here the spin correlations in the x-y plane display algebraic decay, while spin correlations in the z-axis exhibit a power-law decay to a finite value. The dynamical critical exponent is $z=1$.
\item{} A saturated ferromagnetic phase, which in the mean-field approximation occurs at $\lambda_2<-2\lambda_4$. Here excitations are gapped.
\end{enumerate}
\subsection{Nahum transition}
The phase transition of interest occurs on the line separating the two Luttinger liquids. Here there is true long-range order in the x-y plane. The dynamical critical exponent $z$, the crossover scale exponent $\nu$, and the scaling dimension $\Delta_\psi$ of the $\psi$-field have been computed to second order in an epsilon expansion around $d=2$ in Ref.~\cite{flores2025evidence}
\begin{align}
z&=2-\frac{2}{243}\epsilon^2\ ,\qquad \Delta_\psi=\frac{1}{2}(3-z)\ ,\nn
\nu&=\frac{1}{2}\big[1+\frac{\epsilon}{6}-\epsilon^2\big(\frac{1}{36}+\frac{2\gamma_1}{27}+\frac{2\gamma_2}{9}\big]\ ,\nn
\gamma_1&=\frac{2}{27}+\frac{1}{4}+3\ln(2)-\frac{3}{2}\ln(3)\ ,\quad
\gamma_2=-\frac{\ln(2)}{2}+\frac{\ln(3)}{4}-\frac{1}{24}-\frac{1}{27}.
\end{align}
Setting $\epsilon=1$ gives
\be
z\approx 1.99177\ ,\quad
\nu\approx 0.558195.
\label{eq:z}
\ee

\subsection{Microscopic realization in a U(1)-symmetric spin-1 chain}
A suitable microscopic Hamiltonian that is expected to display the transition and proximate phases is the spin-1 Heisenberg ferromagnet with interactions that break the O(3) symmetry down to U(1) \cite{nahum2025continuous}
\begin{equation}
H = -J \sum_{j} {\bf S}_j \cdot {\bf S}_{j+1} +g_1\sum_{j}  (S^z_j)^2+g_2\sum_{j}  (S^z_j)^2(S^z_{j+1})^2\ .
\label{eq:Hamiltonian}
\end{equation}
The field theory parameters $\lambda_{2,4}$ are functions of $g_{1,2}$ as long as the latter are small.

In the following we present results of extensive density matrix renormalization group (DMRG)\cite{dmrg1,dmrg2} simulations of the ground states and low-lying excited states of \fr{eq:Hamiltonian} for a range of the parameters $g_{1,2}$.

\section{Numerically determined ground state phase diagram}
\begin{figure}[h!]
    \centering
    \includegraphics[width=0.7\columnwidth]{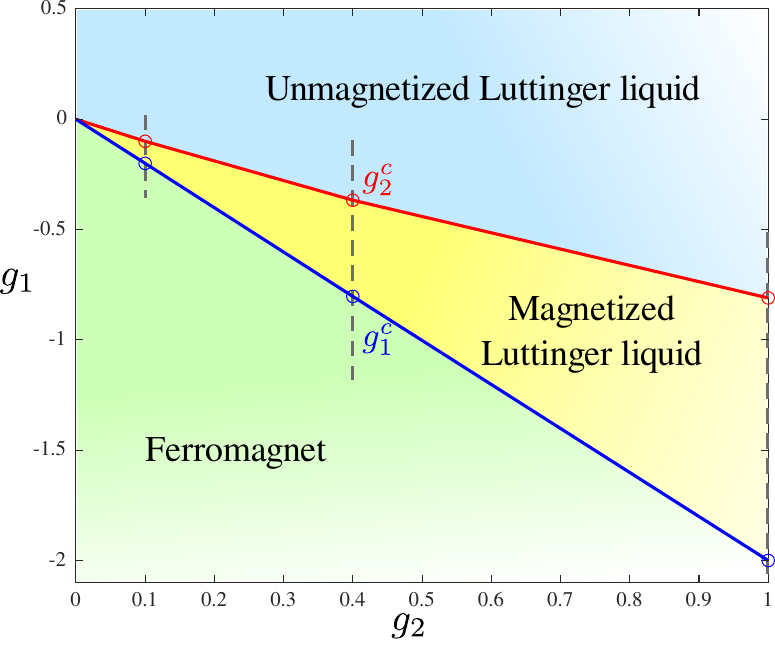}
    \caption{Numerically obtained ground-state phase diagram of the ferromagnetic spin-1 chain with a single-ion anisotropy $g_1$ and a biquadratic coupling $g_2$. The Nahum transition between the unmagnetized and magnetized Luttinger liquids (red line, $g_2^c$)  is the main focus of this paper. The magnetized Luttinger-liquid phase is separated from a saturated ferromagnetic phase by a Japaridze-Nersesyan-Pokrovsky-Talapov transition 
\cite{JaparidzeNersesyan1978,PokrovskyTalapov1979} (blue line, $g_1^c$). We probe the nature of the transitions numerically along vertical cuts (dashed gray lines) at fixed values of $g_2=0.1, 0.4,1$. }
    \label{fig:PD_DMRG}
\end{figure}

In Fig.~\ref{fig:PD_DMRG} we show the ground-state phase diagram of the Hamiltonian \fr{eq:Hamiltonian} obtained from DMRG computations.
We observe precisely the three phases predicted by field theory considerations reviewed above. The simplest phase is the ferromagnetic one: here the ground states are the two saturated ferromagnetic states along the $\pm z$ direction in spin space. In the thermodynamic limit the $\mathbb{Z}_2$ symmetry of spin-flip along the z-direction is spontaneously broken and a unique ground state is selected. The identification of the other phases from DMRG computations, and the determination of the phase boundaries, is addressed in sections \ref{sec:unmagLL}, \ref{sec:magLL} and \ref{sec:NT}. Throughout we proceed by varying $g_1$ for fixed values of $g_2$. This leads to two critical interaction strengths
$g_{1,2}^c(g_2)$ and the following structure of ground-state phases:
\begin{itemize}
\item{} $g_1<g_1^c$: saturated ferromagnet.
\item{} $g_1^c<g_1<g_2^c$: magnetized Luttinger liquid.
\item{} $g_2^c<g_1$: unmagnetized Luttinger liquid.
\end{itemize}
For numerical simulations we use the variational matrix product states (MPS) formulation of two-site DMRG\cite{dmrg1,dmrg2,dmrg3,dmrg4}. Our code has implemented U(1) symmetry thus simulations are restricted to a specified sector of total magnetization $S^z_\mathrm{tot}$. We discard singular values below $10^{-12}$ and keep up to 2000 states; we typically perform up to eight DMRG sweeps and allow up to 200 Lanczos iterations. This allows us to reach the convergence in the critical phases for chains with open boundary condition (OBC) with up to $N=1500$ sites and with up to $N=240$ sites with periodic (PBC) ones.

\section{Unmagnetized Luttinger liquid \sfix{$g_1>g_2^c$}}
\label{sec:unmagLL}
This corresponds to $\lambda_2>0$ in \fr{Lagr} as the effective potential $V(\psi)$ has its minimum at $\psi=0$.

Integrating out the massive $\psi$-field in a mean-field approximation (which is justified as long as $\lambda_2$ is sufficiently larger than zero) and going over to real time we obtain an effective Hamiltonian of Luttinger-liquid form
\be
\mathcal{H}=\frac{v}{2\pi}\int dx \left[K(\partial_x\theta)^2+\frac{1}{K}(\partial_x\phi)^2\right]\ ,\quad
K=\frac{\pi}{\sqrt{2\lambda_2}}+\dots\ .
\label{HLLu}
\ee
Here we have defined the dual field $\phi$ by
\be
\Pi(x)=\frac{1}{\pi}\partial_x\phi(x)\ ,
\ee
where $\Pi(x)$ is the field conjugate to $\theta(x)$. If we approach the Nahum transition (red line in Fig.~\ref{fig:xy03}) by varying $g_1$ in the lattice model at fixed $g_2$ we have
\be
\lambda_2\propto g_1-g_2^c\ .
\ee
This gives the mean-field prediction that close to the transition the Luttinger parameter diverges as
\be
K\propto (g_1-g_2^c)^{-\frac{1}{2}}\ .
\label{Kdiv}
\ee
We expect the exponent to be modified by fluctuations to
\be
K\propto |g_1-g_2^c|^{\nu(1-z)}\ ,
\label{Kmod}
\ee
where the exponents $\nu$ and $z$ are reported in \fr{eq:z}. This gives
\be
\nu(1-z)\approx -0.553601.
\ee
The rationale for \fr{Kmod} is as follows. On the transition low-lying excitations have a dispersion $\omega(k)\sim k^z$, while in the proximate Luttinger-liquid phases the dispersion is linear $\omega(k)=v|k|$. These behaviours can be described by a crossover function
\be
\omega(k)\sim \xi^zf(k\xi)\ ,
\ee
where $\xi\sim |g_1-g_2^c|^\nu$ is the crossover scale and
\be
f(q)\propto\begin{cases}
q & \text{if } q\ll 1\ ,\\
q^z& \text{if } q\gg 1\ .
\end{cases}
\ee
This suggests that the velocity in the Luttinger liquid scales as
\be
v\propto \xi^{1-z}\ .
\ee
Finally, assuming that the stiffness $vK$ remains finite as we approach the transition from the Luttinger-liquid phase we arrive at eqn \fr{Kmod}. This divergence has a direct implication for the transverse correlations: since $C^{xy}(r)\sim r^{-1/(2K)}$ in the adjacent Luttinger liquids, $K\to\infty$ causes the decay exponent to vanish at the transition. This is the Luttinger-liquid manifestation of the true long-range order in the $xy$ plane predicted along the Nahum critical line.
\subsection{Entanglement entropy}
\begin{figure}[ht]
    \centering
    \includegraphics[width=0.5\columnwidth]{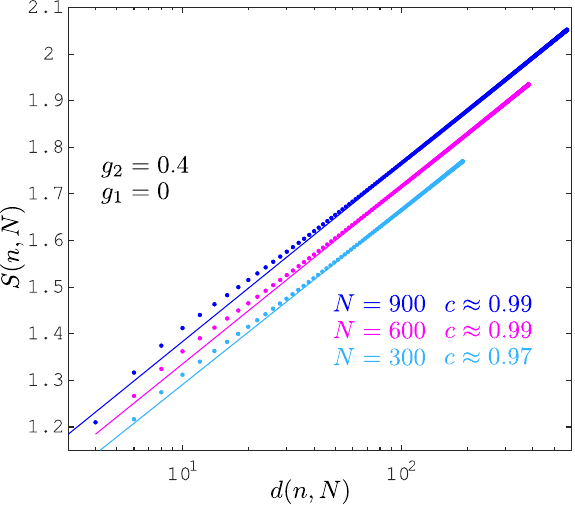}
    \caption{Scaling of the bipartite entanglement entropy $S(n,N)$ in the unmagnetized Luttinger-liquid phase at $g_1=0$, $g_2=0.4$ with OBC and system sizes $N=300,600,900$. The central charge extracted by fitting the DMRG data (dots) to the Calabrese-Cardy formula in Eq.\ref{eq: Calabrese-Cardy} (lines) is in excellent agreement with the Luttinger-liquid prediction $c=1$.}
    \label{fig:centralcharge}
\end{figure}

The Luttinger liquid \fr{HLLu} is a conformal field theory with central charge $c=1$. The latter can be determined from the bipartite entanglement entropy (BEE)\cite{calabreseEntanglementEntropyQuantum2004}. For a subsystem of size $n$ in a finite chain of length $N$ the latter is given by
\begin{align}
\label{eq: Calabrese-Cardy}
S(n,N) &= \frac{c}{6} \ln d(n,N) + s_1 + \ln g,\qquad  d(n,N) =\frac{2N}{\pi}\sin\left( \frac{\pi n}{N}\right)\qquad  \text{OBC};\\
\label{eq: Calabrese-Cardy_pbc}
S(n,N) &= \frac{c}{3} \ln d(n,N) + s_1,\qquad \hskip1.1cm  d(n,N) =\frac{N}{\pi}\sin\left( \frac{\pi n}{N}\right)\qquad  \hskip0.2cm\text{PBC}.
\end{align}
Here $d(n,N)$ is the conformal distance, $\ln g$ the boundary entropy and $s_1$ a non-universal constant. In the open boundary case the interval of length $n$ is taken to start at one of the boundaries. In order to check that the phase at $g_1>g_2^c$ in the spin-1 chain model is indeed a Luttinger liquid we have computed the BEE for open boundary conditions (OBC). For sufficiently large values of the conformal distance these should agree with the CFT prediction \fr{eq: Calabrese-Cardy}.
In Fig.~\ref{fig:centralcharge} we show the entanglement entropy as a function of the conformal distance for $g_2=0.4$ and $g_1=0$ for three different system sizes $N$. For large values of $d(n,N)$ we recover a linear dependence, from which we can extract the central charge that is in very good agreement with the Luttinger-liquid value $c=1$.

\subsection{Spin correlations}
\begin{figure}[h!]
    \centering
    \includegraphics[width=0.47\columnwidth]{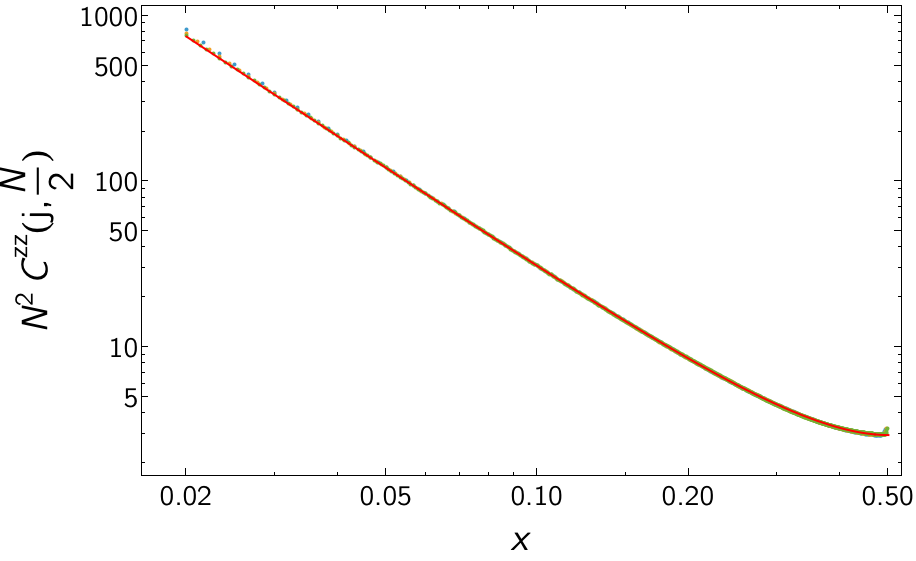}\qquad
     \includegraphics[width=0.47\columnwidth]{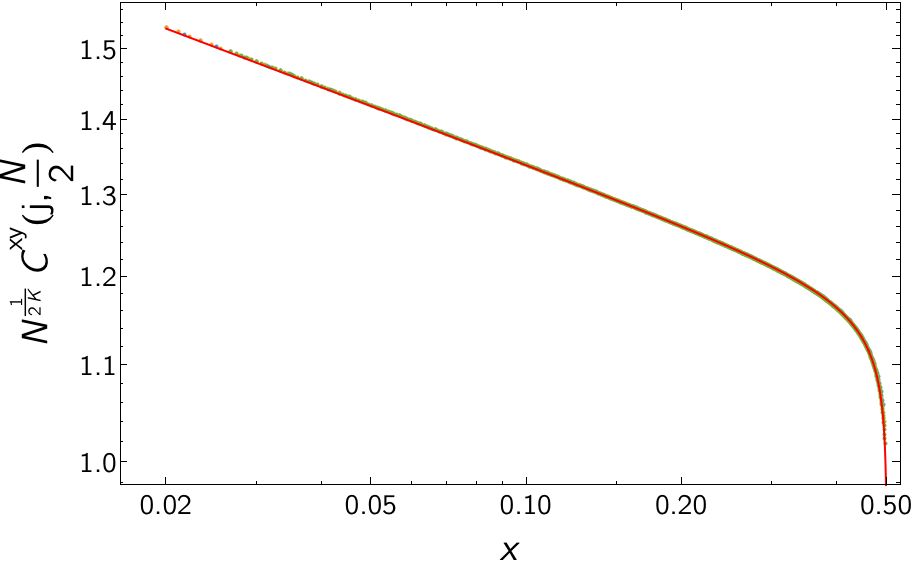}
    \caption{Finite-size scaling collapse for longitudinal and transverse spin correlation functions in the unmagnetized Luttinger-liquid phase for $g_1=0$, $g_2=0.4$ and $N=600,900,1500$. The Luttinger-liquid scaling functions with $K=5.98525$ are shown as the red lines.}
    \label{fig:zz0}
\end{figure}

The long-distance behaviour of spin correlations in the Luttinger liquid phase can be determined by noting that the lattice spin operators are related to the Bose fields by
\begin{align}
S^+_j\sim e^{i\theta(x)}\ ,\quad
S^z_j\sim \frac{1}{\pi}\partial_x\phi(x)\ ,\quad x=ja_0.
\end{align}
We can then use the results of Ref.~\cite{cazalilla2004bosonizing} for open boundary conditions to conclude that
\begin{align}
C^{zz}(j,\frac{N}{2})&=-\langle S^z_{\frac{N}{2}}S^z_{\frac{N}{2}+j}\rangle_c
\simeq \frac{K}{2\pi^2}\left[\frac{1}{d^2(j,2N)}+\frac{1}{d^2(j+N,2N)}\right]+\dots\ ,\nn
C^{xy}(j,\frac{N}{2})&=\langle S^+_{\frac{N}{2}}S^-_{\frac{N}{2}+j}\rangle_c\simeq A_{+-}\left[\frac{d(2j,2N)d(N,2N)}{d(j+\frac{N}{2},2N)d(j-\frac{N}{2},2N)}\right]^{\frac{1}{2K}}+\dots\ ,
\end{align}
where $A_{+-}$ is a non-universal amplitude, the dots indicate sub-leading terms and
\be
d(j,N)=\frac{N}{\pi}\sin\big(\frac{\pi j}{N}\big)\ .
\ee
In order to test the validity of these field theory predictions we check the finite-size scaling collapse of the numerical data and compare it to the Luttinger-liquid scaling functions. Defining rescaled variables
\be
x=\frac{j}{N}\ ,
\ee
we have e.g.
\be
N^2C^{zz}(j,\frac{N}{2})\simeq\frac{K}{2\pi^2}\left[\frac{1}{\bar{d}^2(x)}+\frac{1}{\bar{d}^2(x+1)}\right]\ ,\quad
\bar{d}(x)=\frac{2}{\pi}\sin\big(\frac{\pi x}{2}\big)
\ee
In Fig.~\ref{fig:zz0} we show finite-size scaling plots of $C^{zz}$ and $C^{xy}$ computed by DMRG for system sizes $N=600,900,1500$ for $g_1=0$, $g_2=0.4$. We observe a highly convincing scaling collapse and excellent agreement with the Luttinger-liquid scaling functions.

\begin{figure}[h!]
    \centering
        \includegraphics[width=0.47\columnwidth]{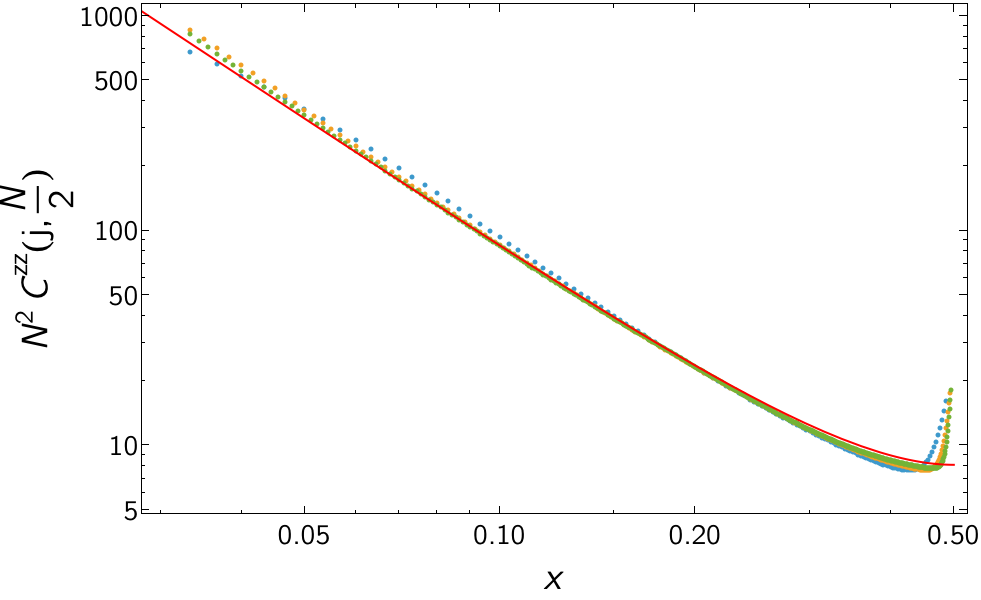}\qquad
    \includegraphics[width=0.47\columnwidth]{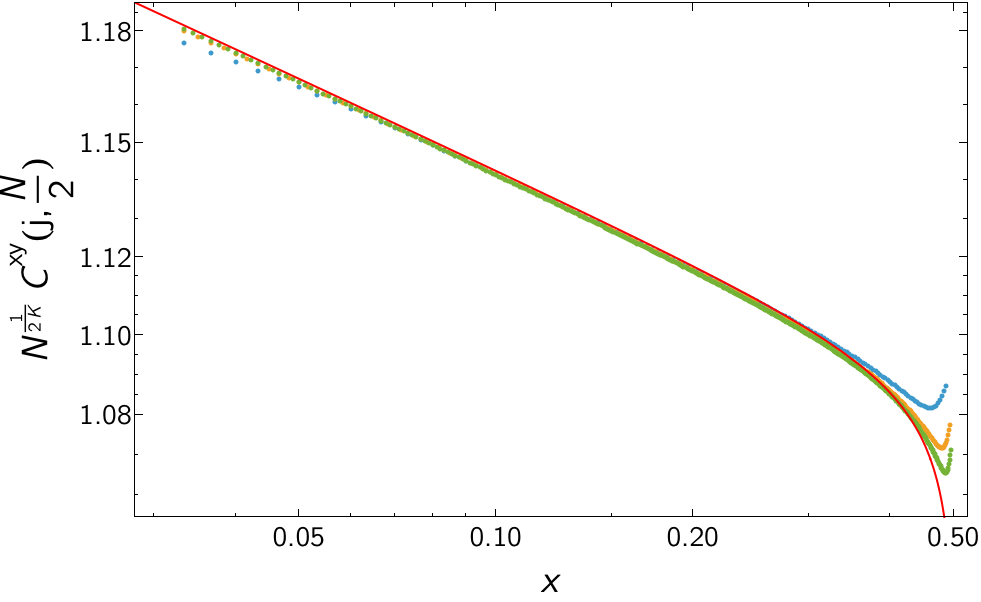}
    \caption{Approximate scaling collapse in the unmagnetized Luttinger-liquid phase for $g_1=-0.3$, $g_2=0.4$ and $N=300,600,900$. The Luttinger-liquid scaling functions with fitted value $K=16.436$ of the Luttinger parameter are shown as the red lines. }
    \label{fig:xy03}
\end{figure}

As we approach the transition the scaling collapse becomes poorer. In Fig.~\ref{fig:xy03} we show the scaling collapse of $C^{zz}$ and $C^{xy}$ computed by DMRG for system sizes $N=600,900,1500$ for $g_2=0.4$, $g_1=-0.3$ (our estimate of the critical point for $g_2=0.4$ is $g_2^c\approx-0.366$). The value of the Luttinger parameter extracted by requiring collapse is $K=16.436$.
We see that for both transverse and longitudinal correlations the scaling collapse becomes poor as we approach the boundary, which suggests that the available system sizes are still too small compared to the short-distance cutoff of the Luttinger liquid.

\section{Magnetized Luttinger liquid \sfix{$g_1^c<g_1<g_2^c$}}
\label{sec:magLL}
This phase corresponds to $\lambda_2<0$ in \fr{Lagr}. Here mean-field theory predicts a finite magnetization along the z-direction
\be
|\psi|\sim\sqrt{-\frac{\lambda_2}{2\lambda_4}}\ .
\label{psiMF}
\ee
There are two degenerate minima connected by the spin-flip symmetry $\psi\rightarrow -\psi$. In the thermodynamic limit this $\mathbb{Z}_2$ symmetry gets spontaneously broken.
For simplicity we will focus on positive magnetizations from here on.
The mean-field result gets significantly modified by fluctuations. The behaviour at small $\lambda_2$ follows from the renormalization-group analyses of Refs~\cite{nahum2025continuous,flores2025evidence}
\be
\psi\sim |\lambda_2|^\beta\ ,\quad
\beta=\nu\Delta_\psi\approx 0.28\ .
\label{eq:betaprediction}
\ee

As sketched in Appendix \ref{app:PT} the transition to the saturated ferromagnet at $\pm\psi\approx 1$ is in the Japaridze-Nersesyan-Pokrovsky-Talapov (JNPT) universality class  \cite{JaparidzeNersesyan1978,PokrovskyTalapov1979}, which gives
\be
1-\psi\sim |\lambda_2-\lambda_{2,c}|^\frac{1}{2}\ .
\ee
These field-theory predictions translate to the lattice model as follows. If we vary $g_1$ at fixed $g_2$ in the lattice model the magnetized Luttinger-liquid phase occurs in the interval
$g_1^c(g_2)<g_1<g_2^c(g_2)$.
Field theory then predicts
\begin{align}
\frac{\langle S^z\rangle}{N}&\propto |g_1-g_2^c|^\beta\ ,\quad g_1\to g_2^c\ ,\nn
1-\frac{\langle S^z\rangle}{N}&\propto \sqrt{g_1-g_1^c}\ ,\quad g_1\to g_1^c\ .
\label{magn}
\end{align}

The derivation of the effective Luttinger-liquid Hamiltonian at low energies proceeds along the same lines as in section~\ref{sec:unmagLL} and we obtain the same divergence of the Luttinger parameter as we approach the phase transition
\be
K\propto |g-g_c|^{\nu(1-z)}\ .
\label{eq:Kdiverge}
\ee

\begin{figure}[ht]
    \centering
    \includegraphics[width=0.9\columnwidth]{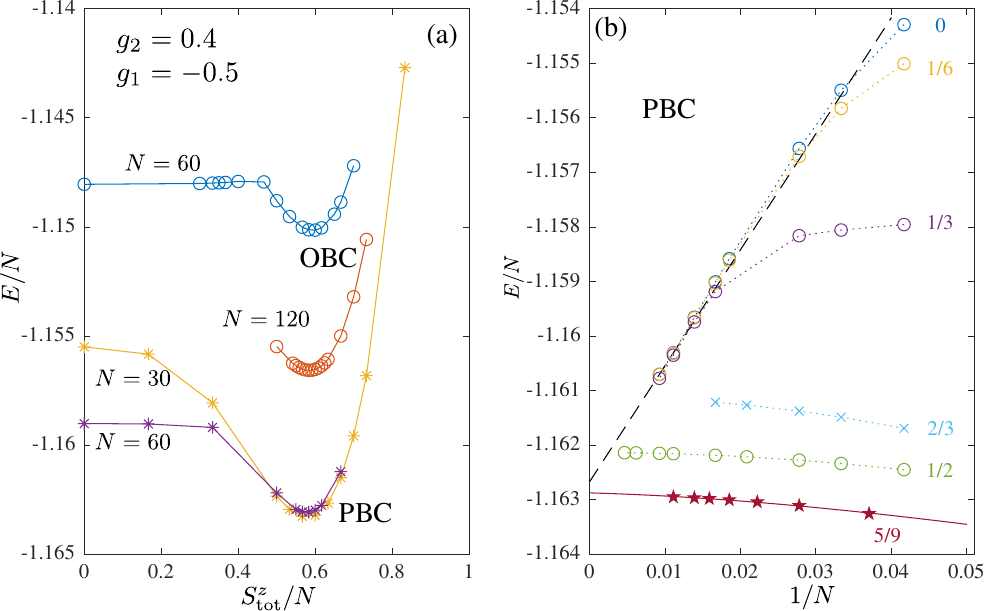}
    \caption{(a) Ground-state energy per site as a function of magnetization for a finite-size chains with open and free (circles) and periodic (stars) boundary conditions. (b) Finite-size scaling of the ground-state energy per site for a fixed magnetization $S^z_\mathrm{tot}/N$ indicated for each data set before (circles), near (stars) and after (crosses) the minimum. }
    \label{fig:Esz}
\end{figure}

In order to determine the ground state numerically we compute the lowest energy as a function of magnetization. Results for both open and periodic boundary conditions at $g_1=-0.5$, $g_2=0.4$ are shown in Fig.~\ref{fig:Esz}.
We observe that the energy per site has a minimum around $S^z_{\rm tot}/N\approx 0.6$ (panel (a)) and exhibits a plateau at small magnetizations (panels (a) and (b)). The plateau value exhibits significant finite-size effects, but as shown in Fig.~\ref{fig:Esz}(b) with increasing system size the ground states in the sectors with small magnetization all approach the same energy per site. We show in subsection \ref{ssec:PS} that the energy difference between the plateau and the minimum extrapolates to a finite value in the thermodynamic limit and that the lowest-energy states in the sectors with small $S^z_\mathrm{tot}/N$ display phase separation.

\begin{figure}[ht]
    \centering
    \includegraphics[width=0.5\columnwidth]{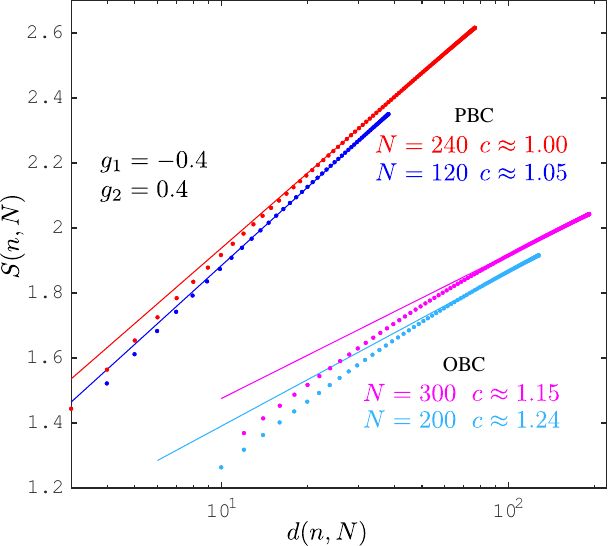}
    \caption{Scaling of the  bipartite entanglement entropy  for systems with OBC with $N=200$, $S^z_\mathrm{tot}=75$ and $N=300$, $S^z_\mathrm{tot}=111$ and PBC with $N=120$, $S^z_\mathrm{tot}=42$ and $N=240$, $S^z_\mathrm{tot}=86$. Each listed magnetization sector corresponds to the finite-size ground state. In each set the data for smaller system sizes are shifted downwards by 0.05 for clarity. We discard from the fit about 20$\%$ (7$\%$) of data points near the edges for OBC (PBC). A value for the central charge is extracted numerically by fitting the DMRG results (dots) to the Calabrese-Cardy formulas for OBC and PBC in Eq.\ref{eq: Calabrese-Cardy}-\ref{eq: Calabrese-Cardy_pbc} (lines). While the results with OBC show significant finite-size effects, the central charge extracted for PBC is in excellent agreement with the Luttinger-liquid prediction $c=1$.
    }
    \label{fig:centralchargeM}
\end{figure}

The overall ground state of the system has a well defined magnetization, which on a finite-size chain can be associated with the minimum of the energy observed in Fig.\ref{fig:Esz}(a). However, boundary effects shift the location of the minimum. This can be seen in Fig.\ref{fig:Esz}(a), where increasing the system size with OBC from $N=60$ to $120$ shifts the average magnetization of the ground state from $S^z_\mathrm{tot}\approx0.6$ to $S^z_\mathrm{tot}\approx0.58$. For periodic boundary conditions this effect is negligible except in the vicinity of the phase transitions.
This finite-size shift of the ground-state magnetization might also explain the deviation of the numerically extracted values of the central charge presented in Fig.\ref{fig:centralchargeM} from the theory predictions. In this respect, simulation of the magnetized Luttinger liquid with PBC seems the most obvious strategy. However, it comes with a price of significantly higher computational cost. For a typical simulation like the one presented in Fig.\ref{fig:centralchargeM} the entanglement for a periodic system with $N=240$ sites is higher than entanglement of a comparable system size with OBC by about $\Delta S(n,N)\approx0.6$. This implies that to keep the same accuracy of the simulations, the DMRG bond dimension has to increase by a factor $\exp(\Delta S)\approx 1.8$ and the complexity of the algorithm ($\propto D^3$) is therefore 6-times larger. In practice it means that we can reach convergence only for relatively small system sizes.

\subsection{Entanglement entropy}
In order to verify the Luttinger-liquid description of the $g_1^c<g_1<g_2^c$ phase we have computed the BEE for both open and periodic boundary conditions. The results extracted for periodic boundary conditions give a convincing agreement with the Luttinger-liquid prediction $c=1$.  In contrast, we observe that finite-size corrections are significantly stronger for OBCs, the agreement with the conformal-field theory prediction is poor, and extracting the central charge therefore would require larger system sizes. A possible explanation for this is that the CFT prediction \fr{eq: Calabrese-Cardy} is based on conformally invariant boundary conditions, and these are approached in the lattice model only for sufficiently large values of $d(n,N)$.

\subsection{Ground state magnetization per site}
\begin{figure}[h!]
    \centering
    \includegraphics[width=0.9\columnwidth]{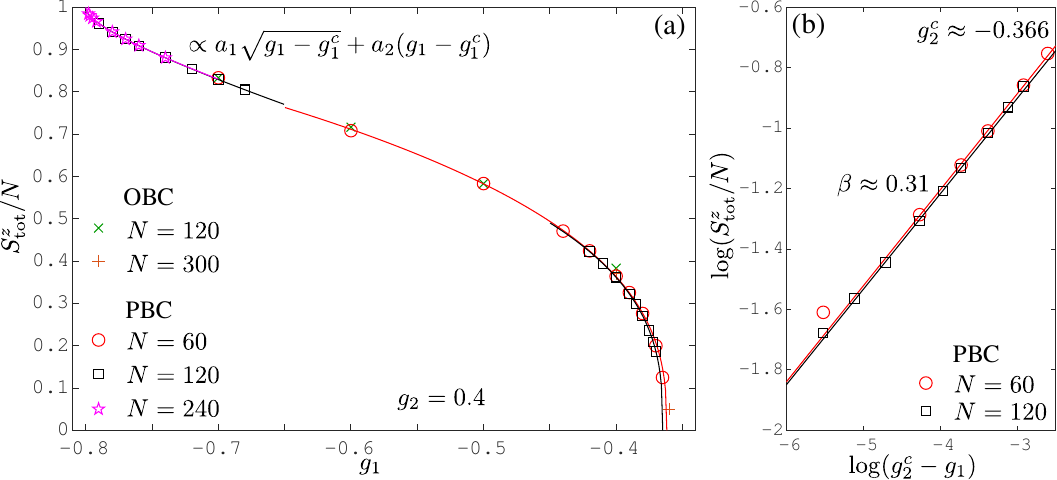}
    \caption{ (a) Ground-state magnetization per site $S^z_\mathrm{tot}/N$ in magnetized Luttinger-liquid  phase for $g_2=0.4$. Near the saturated ferromagnetic phase the scaling is compatible with a JNPT transition in Eq.\ref{eq:PTscaling} (black and magenta lines). Upon approaching the Nahum transition the magnetization scales to zero with critical exponent $\beta\approx0.31$, which agrees within $10\%$ with the field theory prediction \ref{eq:betaprediction}. (b) Data near the Nahum transition on a log-log scale. We have fixed the location of the transition to $g_2^c\approx-0.366$ as determined in section~\ref{ssec:locate}. In the linear fits we use all shown data points except the first and last ones.
          }
    \label{fig:Sz}
\end{figure}

In Fig.~\ref{fig:Sz} we show the ground-state magnetization per site as a function of $g_1$ at fixed $g_2=0.4$. We restrict our analysis to positive values of $S^z_{\rm tot}$, in the understanding that for finite systems there is a $\mathbb{Z}_2$ spin-flip symmetry that connects the sectors with $\pm S^z_{\rm tot}$.
For each point we numerically identify the magnetization sector of the ground state $S^z_\mathrm{tot}$. Then, in order to reduce the error from the finite resolution of the magnetization sectors ($\Delta(S^z_\mathrm{tot}/N)=1/N$), we perform spline fits over 5 points near the ground state (at this stage we already ensure the highest possible resolution with $S^z_\mathrm{tot}$ changing in steps of 1 near the minimum). We associate the finite-size magnetization with the minimum of the spline curve near the ground state.  For comparison we also show a few results with OBC.

\begin{figure}[h!]
    \centering
    \includegraphics[width=0.9\columnwidth]{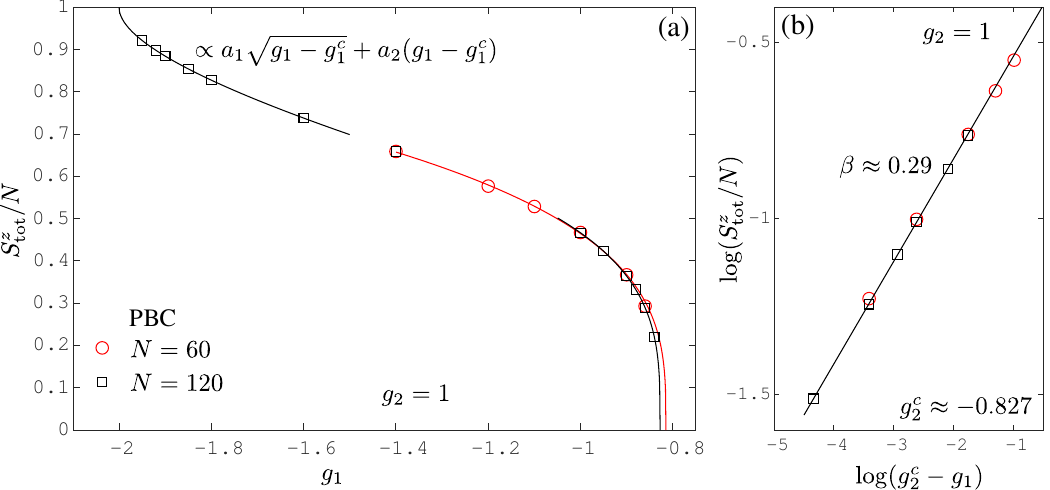}
    \caption{ Same as Fig.\ref{fig:Sz} but for $g_2=1$. The numerically extracted critical exponent $\beta\approx 0.29$ is quite close to the field theory prediction $\beta\approx0.28$.
          }
    \label{fig:Sz_g1}
\end{figure}

In Fig.\ref{fig:Sz}(b) we show the scaling of the magnetization with the distance to the Nahum transition on a log-log scale. The slope gives $\beta\approx0.31$, which is approximately $10\%$ higher than the field theory prediction in (\ref{eq:betaprediction}). This discrepancy can be attributed in part to the proximity to the multicritical SU(2)-invariant point at $g_1=g_2=0$. If we move further away from this point the agreement is significantly improved. The example $g_2=1$ is shown in Fig.\ref{fig:Sz_g1}. Here we find $\beta\approx0.29$, which is quite close to the value $\beta\approx 0.28$ predicted by the epsilon expansion.
However, it is important to note that due to slower convergence we cannot approach the critical point as closely as we did for $g_2=0.4$.
Our numerical results for the scaling of the magnetization per site in the vicinity of the transition to the saturated ferromagnetic phase shown in Fig.\ref{fig:Sz}(a) are in excellent agreement with the field-theory predictions \fr{magn} if we allow for subleading corrections
\be
\frac{S^z_{\rm tot}}{N}\approx 1-a_1\sqrt{g_1-g^c_1}+a_2(g_1-g^c_1)\ .
\label{eq:PTscaling}
\ee
The subleading term is motivated by the low-density expansion of a generic 1D dilute gas with a finite two-particle scattering length.

\subsection{Spin-correlations }
\begin{figure}[h!]
    \centering
    \includegraphics[width=0.47\columnwidth]{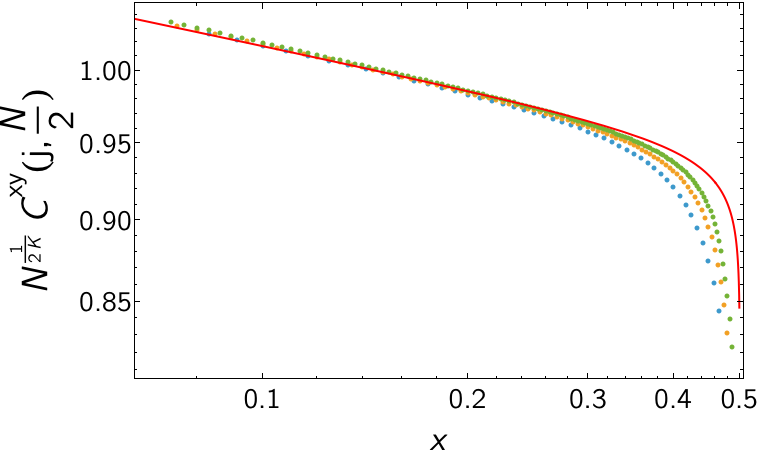}\qquad
     \includegraphics[width=0.47\columnwidth]{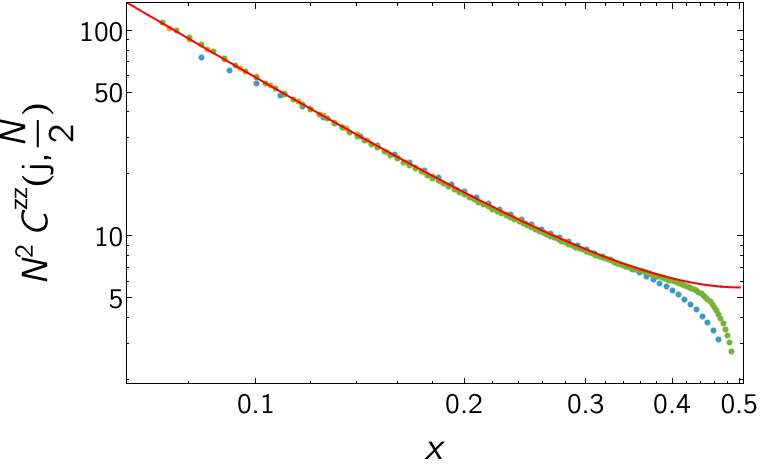}
    \caption{Approximate scaling collapse for $g_1=-0.4$, $g_2=0.4$ and $N=120,200,300$. The Luttinger parameter is taken to be $K=11.3805$. }
    \label{fig:xy04}
\end{figure}

In order to verify the prediction that the low-energy physics is described by a Luttinger liquid we have computed connected spin-spin correlation functions.
In Fig.~\ref{fig:xy04} we show the approximate finite-size scaling collapse of transverse and connected longitudinal spin correlations for $g_1=-0.4$, $g_2=0.4$ and system sizes $N=120,200,300$ (with periodic boundary conditions).
The DMRG results are compared to the Luttinger-liquid scaling functions where the Luttinger parameter $K=11.3805$ has been fixed by performing a fit of $N^2C^{zz}(j,N/2)$ to the Luttinger-liquid scaling function. We see that the scaling collapse is good, given the rather modest system sizes available.

\subsection{Phase-separated states}
\label{ssec:PS}
We now turn to a discussion of phase-separated states, which we will show to occur at a finite energy (in the thermodynamic limit) above the ground state. Understanding these states is important in order
to establish the energy scale below which the Luttinger-liquid description of the partially magnetized critical phase applies.

\begin{figure}[ht]
    \centering
    \includegraphics[width=0.95\columnwidth]{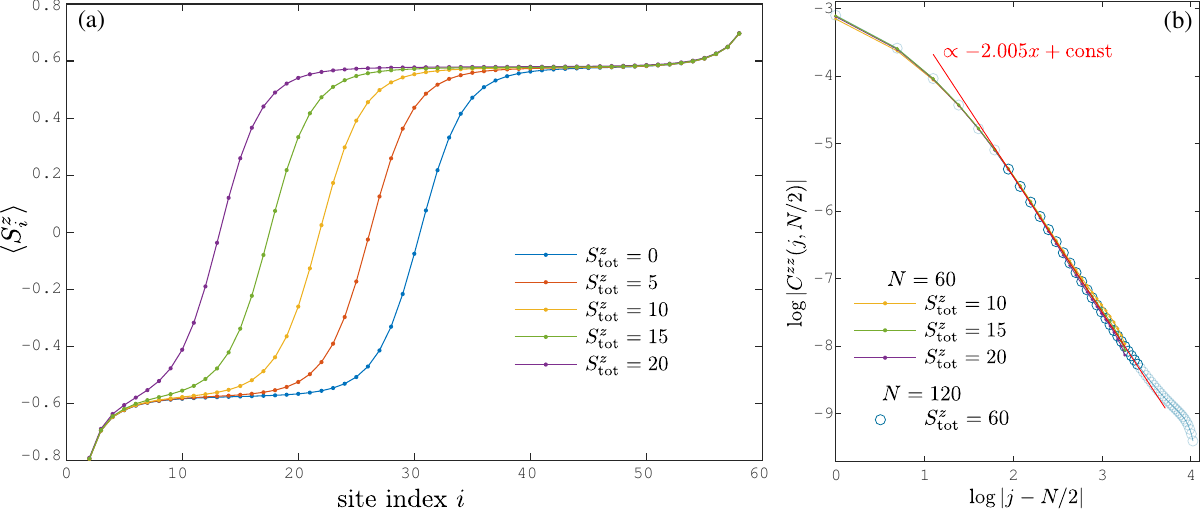}
    \caption{ (a) Phase separation in system with boundaries polarized in the opposite direction. The shown example is for  $N=60$, $g_1=-0.5$ and $g_2=0.4$ and pinning boundary fields $h^z_{L,R}= \pm 1$.       Upon changing the total magnetization the DW shifts, but shows no noticeable change in its shape. (b) Scaling of the correlation function computed within the largest plateaux for two different system sizes. The slope in log-log scale is in excellent agreement with field theory prediction $-2$. The points used in the fit are shown in dark open circles. }
    \label{fig:DW}
\end{figure}

One of the difficulties in determining the physical properties in the magnetized Luttinger-liquid phase is the tendency to phase-separate in the low-lying excitations. One way to see this is to work with open boundary conditions, where the magnetization profile in the ground state obtained by DMRG exhibits ferromagnetic domain walls (DWs). By introducing the opposite boundary field at the edges we force ferromagnetic domains with positive and negative magnetization near the selected boundary, and thus induce the appearance of a single DW. Importantly, the boundary fields required to achieve this are fixed and independent of system size.
The relative extent of the domains is determined by the total magnetization $S^z_{\rm tot}$ as shown in Fig.\ref{fig:DW}(a). We observe that the bulk magnetization within each ferromagnetic domain approximately takes the values of the two Luttinger-liquid ground states connected by spin-flip $\mathbb{Z}_2$ symmetry. Importantly, neither the bulk magnetization within each ferromagnetic domain nor the shape of the DWs show noticeable changes upon tuning the magnetization $S^z_\mathrm{tot}$, which is fully consistent with the phase separation picture. 
For periodic boundary conditions we find phase-separated states with a finite gap to the ground state (see below). 

In order to set the stage for a discussion of the phase-separated states it is useful to recall how phase separation manifests itself in the exactly solvable spin-1/2 XXZ chain in the ferromagnetic regime
\be
H_{\rm XXZ}=-\sum_{j=1}^N\big(\sigma^x_j\sigma^x_{j+1}+\sigma^y_j\sigma^y_{j+1}+\Delta\sigma^z_j\sigma^z_{j+1}\big)\ ,\quad \Delta>1.
\ee
Here the absolute ground states are the two saturated ferromagnetic states polarized along the z-direction. In contrast, the ground states in the sectors with finite magnetization are phase-separated states that in terms of the Bethe Ansatz solution correspond to multi-particle bound states \cite{gaudin2014bethe,albertini1995xxz}.
The ground-state energy for large $N$ in the sectors with magnetization per site $m$ takes the form
\be
E(m)=-N\Delta+4\sinh\gamma+{\cal O}(e^{-\alpha N})\ ,\quad \Delta=\cosh\gamma.
\label{EPS}
\ee
The physical interpretation of \fr{EPS} is as follows. The leading term is the energy of the absolute ground states (the two saturated ferromagnetic states along the $\pm$z-direction). The second term is the finite energy cost for forming two DWs (for periodic boundary conditions) between ferromagnetic domains polarized along the $\pm z$ direction in spin space. The finite-size corrections are exponentially small in system size.
In each magnetization sector there are $N$ states whose energies are exponentially close in $N$ to $E(m)$. They are labeled by different momentum eigenvalues. The energy cost required for the minority domain to move with a finite momentum is exponentially small in system size as the entire bound state has to move as a whole. At large $\Delta$ this requires an extensive order in perturbation theory in the hopping.

The situation here differs from the one in the spin-1/2 XXZ chain in the structure of the domains. The fact that the magnetization of the two domains is related to that of the Luttinger-liquid ground states suggests that if we pin the DW by applying boundary conditions as in Fig.~\ref{fig:DW}, the low-energy properties are described in terms of an inhomogeneous Luttinger liquid of the form
\be
H=\frac{1}{2\pi}\int dx \left[v(x)K(x) (\partial_x\theta)^2+\frac{v(x)}{K(x)}(\partial_x\phi)^2\right]\ .
\ee
Here the velocity $v(x)$ and Luttinger parameter $K(x)$ are position-dependent, and approximately reduce to the values in the Luttinger-liquid ground state sufficiently far away from the position of the DW. This picture is supported by DMRG results for the spin-correlation function $C^{zz}(j,N/2)$ shown in Fig.~\ref{EPS}(b), which is seen to display a $x^{-2}$ power-law decay. The observed behaviour is similar to the phase separation reported recently in the nematic Luttinger liquid that appears in XXZ chain in the presence of the next-nearest-neighbor Ising interaction\cite{phase_separation}. There, similar to the present case, the domain wall separating two Luttinger liquids with positive and negative magnetization appears to be well localized. 

For periodic boundary conditions a field-theory description is less obvious. One possibility is to introduce the centre-of-mass coordinate $X$ of the minority domain as an independent variable that parametrizes the profile of an inhomogeneous Luttinger liquid, and to add a kinetic term $P_X^2/2M$ to the Hamiltonian. This would give a translationally invariant theory appropriate for periodic boundary conditions. Deriving such a theory is beyond the scope of this paper, but some insight can be gained from the energies of the phase-separated states shown in Fig.~\ref{fig:Esz}(b), which for the system sizes considered correspond to magnetizations $1/3$, $1/6$, and $0$. We see that the scaling of energies with system size is given by
\begin{align}
\frac{E_{\rm PS}(m)}{N}&=e_0+\frac{a_1}{N}+o(N^{-1}),
\label{EPS2}
\end{align}
where $e_0$ is the energy per site of the Luttinger-liquid ground state. The accuracy of our DMRG data does not allow us to deduce the higher order corrections in \fr{EPS2}. The leading correction in \fr{EPS2} can be interpreted as a finite energy cost of two independent DWs. The fact that the deviations from the $N^{-1}$ scaling increase with larger values of the magnetization $S^z_{\rm tot}/N$ indicates that the DWs start to interact due to their reduced separation for these system sizes. Indeed, assuming that the structure of DWs for PBCs is analogous to the one shown in Fig.~\ref{fig:DW} for non-zero boundary fields, i.e. each DW interpolates between the Luttinger-liquid ground-state values $\pm m_{\rm GS}$, increasing $S^z_{\rm tot}/N$ reduces the separation of the two DWs and hence increases their interaction energy. 

We take the results discussed above as confirmation of the phase-separated nature of the ground states in the sectors with fixed $S^z_{\rm tot}>m_{\rm GS}$ for sufficiently large $N$. Interestingly, as in the case of the spin-1/2 XXZ chain, the phase separated states have a finite excitation gap over the Luttinger-liquid ground state in the thermodynamic limit.

\section{Vicinity of the Nahum transition}
\label{sec:NT}
In order to determine the exponents $z$ and $\nu$ we need to locate the transition in parameter space. To do so we vary $g_1$ at fixed $g_2$ in the lattice model. One prediction of the Luttinger-liquid theories of the proximate phases is that their respective velocities $v$ vanish and their Luttinger parameters $K$ diverge, while $vK$ tends to a constant as we approach the transition.
\subsection{Locating the transition}
\label{ssec:locate}
\begin{figure}[h!]
    \centering
    \includegraphics[width=0.99\columnwidth]{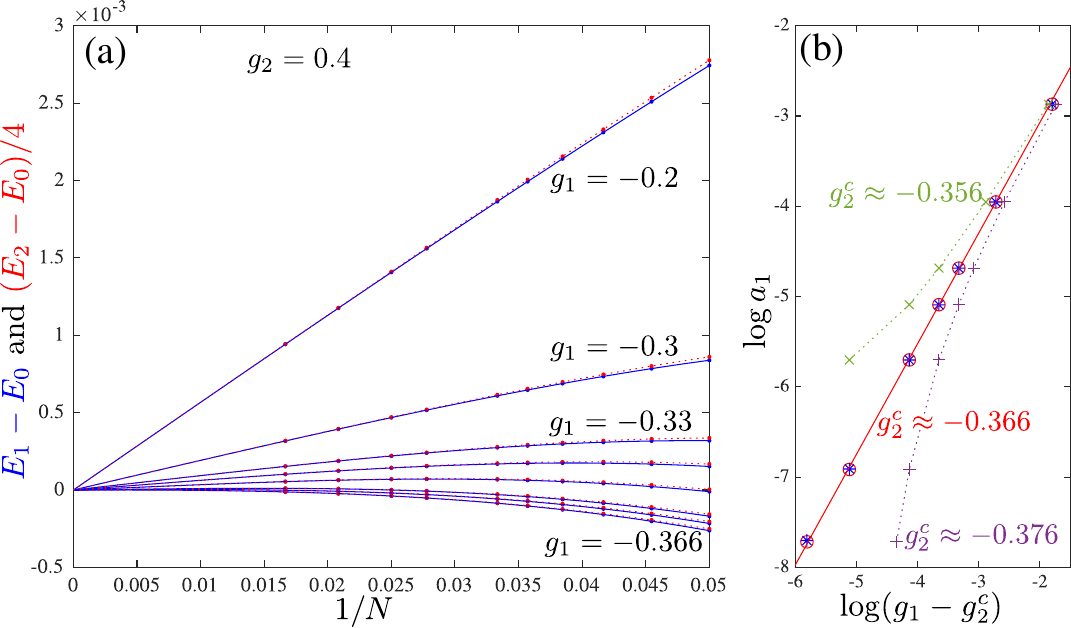}
    \caption{ (a) Finite-size scaling of the excitation energy between the ground state in the sector with $S^z_\mathrm{tot}=0$ and the lowest energy states $E_1$ and $E_2$ in the sectors with $S^z_\mathrm{tot}=1$ (blue) and $2$ (red) correspondingly.  The results are shown for $g_1=-0.2,-0.3,-0.33,-0.34,-0.35,-0.36,-0.363,-0.366$. Dots are DMRG data, blue solid and red dotted lines are the fit to \fr{EofN}. (b) Scaling of the linear term $a_1$ extracted in (a) with the distance to the transition in the log-log scale. We associate the  separatrix (red line) with the location of the critical point $g_2^c\approx 0.366$; its slope $\frac{\log a_1}{\log(g_1-g_2^c)}\approx 1.21$ agrees within $10\%$ with the prediction in Eq.\ref{eq:magslope} with $2\nu(z-1)\approx1.1072$. For comparison, we present the scaling that assumes the location of the critical point by $\pm 0.01$ away from this value (pluses and crosses) with a noticeable curvature.
          }
    \label{fig:magnetic_gap}
\end{figure}

In the Luttinger-liquid phases close to the transition the finite-size low-energy spectrum for open boundary conditions is given by
\be
E-E_0=\frac{2\pi v}{N}\left[\frac{(\delta S^z)^2}{4K}+N\right]+\mathcal{O}(N^{-2})\ ,
\label{FSS}
\ee
where $\delta S^z$ and $N$ are integers. As $K$ diverges as we approach the transition from the unmagnetized Luttinger-liquid phase the leading corrections to \fr{FSS} are due to band-curvature and other irrelevant operators with integer scaling dimension. Hence the lowest  two energy gaps are expected to have the following expansions
\begin{align}
E_m-E_0&=\frac{2\pi v m^2}{4KN}+\frac{a_2^{(m)}}{N^2}+\frac{a_3^{(m)}}{N^3}+\frac{a_4^{(m)}}{N^4}+\dots\ ,\quad m=1,2.
\label{EofN}
\end{align}
In Fig.\ref{fig:magnetic_gap}(a) we show how the lowest two energy gaps scale with $N^{-1}$ for a range of values of $g_1$. The blue lines are fits to the functional form \fr{EofN}. The coefficient of the leading term vanishes as as power law  as $g_1$ approaches the transition
\be
a_1=\frac{2\pi v }{4K}\propto (g_1-g_1^c)^{2\nu(z-1)}\ .
\label{eq:magslope}
\ee
In Fig.\ref{fig:magnetic_gap}(b) we show the scaling of
$\ln(g_1-g_2^c)$ for several trial values of $g_2^c$. The critical value $g_2^c$ can then be identified by requiring power-law scaling, i.e. a straight line in Fig.~\ref{fig:magnetic_gap}(b). These considerations allow us to determine the critical point with good accuracy. 

\subsection{Divergence of the Luttinger parameter}
\begin{figure}[ht]
    \centering
    \includegraphics[width=0.99\columnwidth]{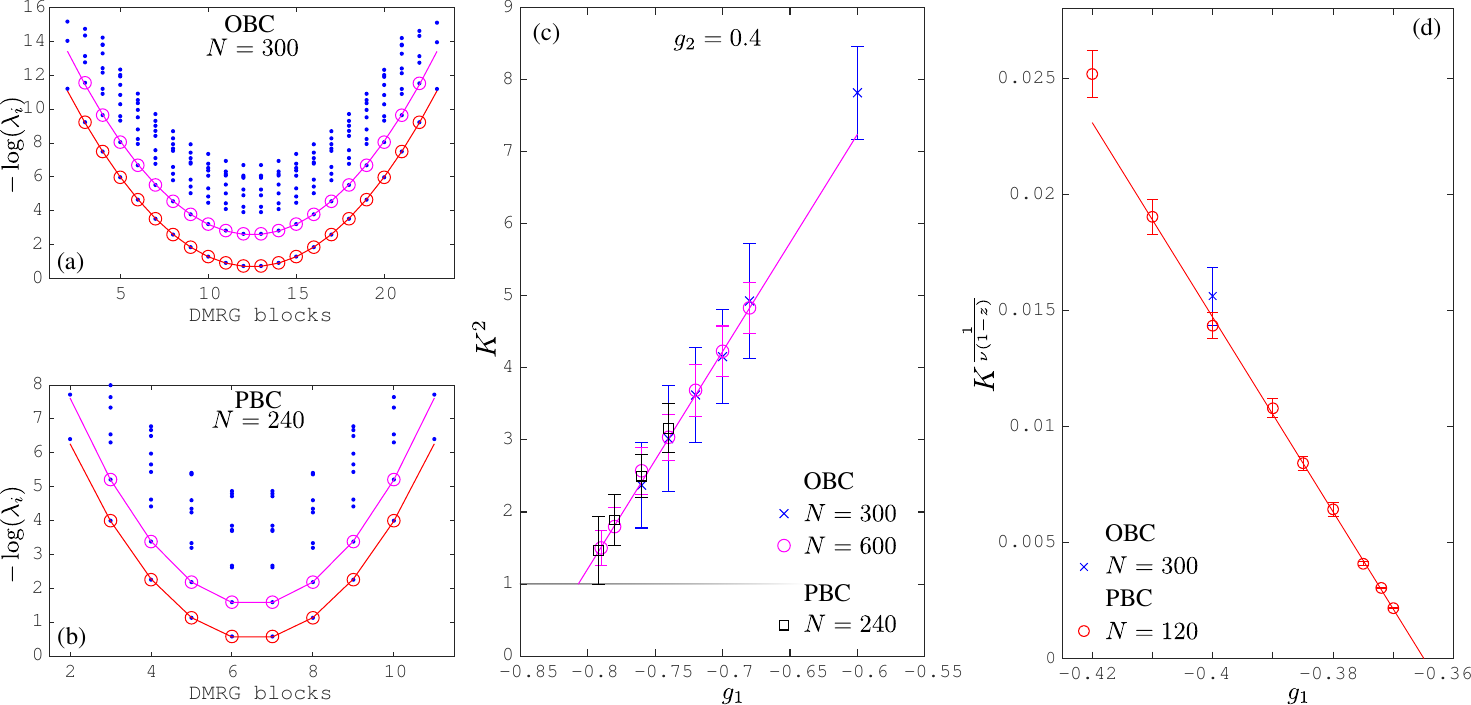}
    \caption{(a)-(b) Examples of the entanglement spectra  extracted from singular values $\lambda_i$ for (a) open and (b) periodic boundary conditions. The spectrum is arranged into DMRG blocks characterized by different half-chain magnetization. The lowest level(s) typically have magnetization $\left[ S^z_\mathrm{tot}/2\right]\pm1$.  (a) Results for $N=300$, $g_2=0.4$, $g_1=-0.4$ and $S^z_\mathrm{tot}=111$. (b) Results for $N=240$, $g_2=0.4$, $g_1=-0.74$ and $S^z_\mathrm{tot}=211$.  The chosen magnetization sectors correspond to the finite-size ground state.   Red and magenta lines are parabola fit of the two lowest envelops $a(x+c)^2+b$; the data points used in the fit are marked with circles. (c)-(d) The scaling of the Luttinger-liquid (LL) parameter extracted from the fit  shown in (a)-(b) as $K=(b_2-b_1)/(2a_1)$, where indices 1 (2) refer to the lowest (second lowest) envelop. The errorbars reflect the difference in numerical values of $K$, if $a_1$ is replaced by $a_2$. (c) Upon approaching the saturated ferromagnet $K$ decays to the free boson value $K=1$ as a square-root resulting in a linear scaling of $K^2$. (d) Upon approaching the transition into unmagnetized LL, the divergence of the LL exponent is consistent with Eq.\ref{eq:Kdiverge} with critical exponents $z$ and $\nu$ as defined in Eq.\ref{eq:z}.}
    \label{fig:estower}
\end{figure}

\begin{figure}[ht]
    \centering
    \includegraphics[width=0.99\columnwidth]{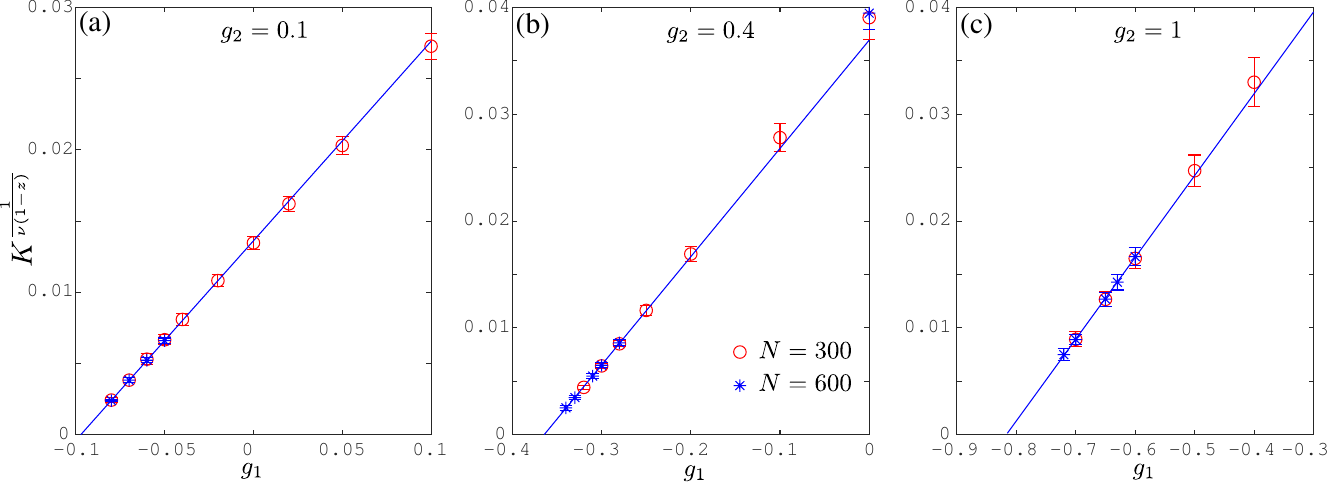}
    \caption{Scaling of the Luttinger-liquid (LL) parameter $K$ with the distance to the transition in the unmagnetized phase for three different values of $g_2$. The divergence of the LL exponent is consistent with Eq.\ref{eq:Kdiverge} with critical exponents $z$ and $\nu$ as defined in Eq.\ref{eq:z}. The values of $K$ were extracted by fitting the two lowest envelops in the entanglement spectrum as shown in Fig.\ref{fig:estower}(a)}
    \label{fig:LLunmagnet}
\end{figure}

Determining the Luttinger parameter as a function of $g_{1,2}$ from the spin-spin correlation functions is numerically quite costly. We therefore proceed as follows.
We compute the entanglement spectrum of the ground state using DMRG. The Luttinger-liquid prediction for open boundary conditions and a subsystem $A$ of length $N/2$ starting at one of the boundaries is \cite{Lauchli2013,LaflorencieRachel2014}
\be
\xi(\delta S^z_A,\{n_k\})=\xi_0+\frac{2\pi^2}{\ln(N/a)}\left(\frac{(\delta S^z_A)^2}{2K}+\sum_k kn_k\right),
\ee
where $a$ is an UV cutoff, $\delta S^z_A$ are the integer differences in total spin in the subsystem and $n_k$ are integers. For periodic boundary conditions we have the same expression with the replacement $2\pi^2\rightarrow \pi^2$. From the
two lowest eigenvalues for a given $\delta S^z_A$ we can then extract the Luttinger parameter as
\be
K=\frac{(\delta S^z_A)^2}{2}
\left[\frac{\xi(\delta S^z_A,n_1=1)-\xi_0}{\xi(\delta S^z_A,n_1=0)-\xi_0}-1\right].
\label{Luttp}
\ee
In Fig.~\ref{fig:estower} we show the $g_1$-dependence of $K$ extracted from \fr{Luttp} for $g_2=0.4$ in the magnetized Luttinger-liquid phase. 
In Fig.~\ref{fig:estower}(d) we show that as $g_1$ approaches the Nahum transition the Luttinger parameter diverges in a way that is compatible with the field theory prediction \fr{Kmod}
\be
K\propto |g_1-g_2^c|^{\nu(1-z)}\ ,\quad \nu(1-z)\approx -0.553601.
\label{KNahum}
\ee
In Fig.~\ref{fig:estower}(c) we show that as $g_1$ approaches the JNPT transition the Luttinger parameter approaches $K=1$ in agreement with eqn \fr{KJNPT}
\be
K-1\propto |g_1-g_1^c|^{\frac{1}{2}}\ .
\ee
Finally, in Fig.~\ref{fig:LLunmagnet} we show the dependence of the Luttinger parameter $K$ on $g_1$ in the unmagnetized Luttinger-liquid phase.
We see that as we approach the transition $K$ diverges in a way that is compatible with the field theory prediction \fr{KNahum} for the Nahum transition.

\subsection{Dynamical critical exponent}
\begin{figure}[h!]
    \centering
    \includegraphics[width=0.99\columnwidth]{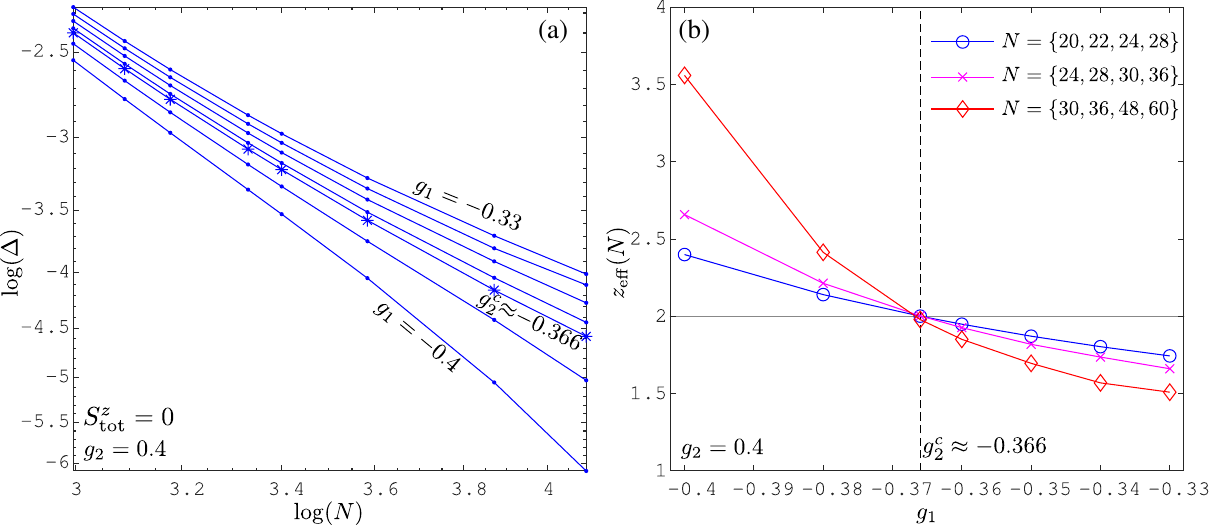}
    \caption{ (a) finite-size scaling of non-magnetic excitation energy extracted by targeting multiple states within $S^z_\mathrm{tot}=0$ sector for $g_2=0.4$ and various values of $g_1$ on both sides of the transition. At the critical point the scaling appears linear in the log-log scale (stars). (b) Effective dynamical critical exponent $z_\mathrm{eff}$ extracted with the linear fit over the indicated sub-sets of system sizes shown in (a). Upon increasing system size, $z_\mathrm{eff}$ asymptotically approach conformal value $z=1$ in the unmagnetized LL, while diverges in the magnetic LL phase, where  $S^z_\mathrm{tot}=0$ is no longer a ground state. The finite-size estimates of $z_\mathrm{eff}$ cross near the transition. At the crossing the value of $z_\mathrm{eff}\approx 2$ which is consistent with theory predictions $z\approx 1.992$.
          }
    \label{fig:dynexp_04}
\end{figure}

In addition, we probe the dynamical critical exponent $z$ from scaling of the excitation energies within the sector of zero magnetization. We extract low-lying excitations along with the ground state by targeting multiple states in DMRG\cite{chepigaExcitationSpectrumDensity2017}. The finite-size scaling near the transition is shown in Fig.\ref{fig:dynexp_04}(a). On a log-log scale the transition appears as a separatrix between convex scaling that asymptotically approaches $z=1$ of the unmagnetized Luttinger liquid and concave scaling when phase separation develops and the zero-magnetization state is no longer critical (and no longer the ground state). We find it instructive to extract an effective dynamical critical exponent $z_\mathrm{eff}$ by fitting sets of points in different ranges of system sizes and tracking the finite-size effects in $z_\mathrm{eff}$ as shown in Fig.\ref{fig:dynexp_04}{b}. As expected, in the non-magnetic Luttinger liquid $z_\mathrm{eff}$ slowly approaches the conformal value of $z=1$. Different finite-size estimates of  $z_\mathrm{eff}$ eventually cross and the crossing appears to happen near the transition and at $z\approx 2$. 
With the available numerical accuracy it is impossible to distinguish the field-theory value $z\approx1.992$ from $z=2$. Our finite-size estimates are therefore consistent with $z\simeq2$ and incompatible with the previously reported numerical estimate $z=1.51\pm0.03$~\cite{flores2025evidence}.

\begin{figure}[h!]
    \centering
    \includegraphics[width=0.99\columnwidth]{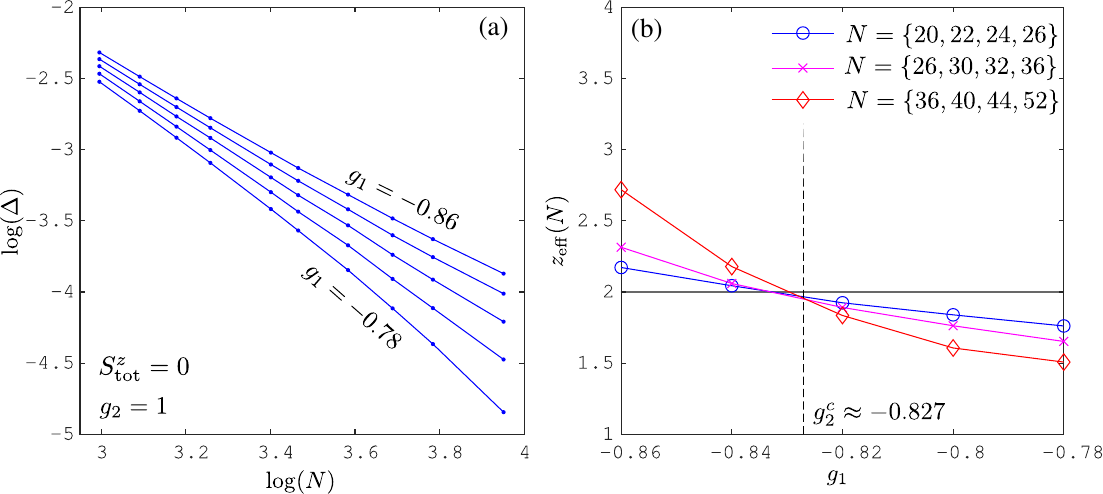}
    \caption{ Same as Fig.\ref{fig:dynexp_04} but for $g_2=1$. At the transition (dashed line) our results are consistent with theory predictions $z_\mathrm{eff}\approx 2$ and disagree with previously reported $z\approx 1.5$\cite{flores2025evidence}.
          }
    \label{fig:dynexp_1}
\end{figure}






\section{Conclusion}
In this work we have carried out extensive numerical simulations of a spin-1 lattice model expected to exhibit the criticality-enabled long-range-order transition proposed by Nahum. Our results are consistent with renormalization-group predictions based on $\epsilon$ expansions~\cite{nahum2025continuous,flores2025evidence}. First, the Luttinger parameter diverges on approaching the transition from the neighbouring Luttinger-liquid phases; since the transverse correlation exponent is proportional to $1/K$, this behavior is consistent with genuine long-range order in the $xy$ plane at the Nahum transition. Second, the exponent $\beta$ governing the vanishing magnetization is close to the $\epsilon$-expansion prediction $\beta_\epsilon\approx0.28$: sufficiently far from the SU(2)-symmetric point, where corrections to scaling are expected, we obtain $\beta_{\rm DMRG}(g_2=1)\approx0.29$. Third, we proposed relation Eq.~\fr{KNahum} which expresses the divergence of the Luttinger parameter to the critical exponents $\nu$ and $z$, and our numerical results are compatible with the corresponding $\epsilon$-expansion prediction. Finally, finite-size scaling of the low-lying spectrum gives estimates consistent with $z\simeq2$; with the present numerical accuracy we cannot distinguish $z=2$ from the $\epsilon$-expansion value $z_\epsilon\approx1.992$~\cite{flores2025evidence}. These results are incompatible with the previously reported estimate $z\approx1.51$ based on the dynamical structure factor~\cite{flores2025evidence}.

An interesting feature of the partially magnetized Luttinger-liquid phase is the presence of phase-separated excitations with a finite gap above the ground state. We have briefly sketched what field theory descriptions of these states may look like, but it would be interesting to derive a theory of these states in the large-$S$ limit.

\section*{Acknowledgements}
We are grateful to Adam Nahum for very helpful discussions, in particular on the phase separated states. This work was supported in part by the EPSRC under grant EP/X030881/1 (FHLE) and by the European Union through the ERC grant TRANGINEER/101220181 (NC). The views and opinions expressed are those of the authors only and do not necessarily reflect those of the European Union or the European Research Council Executive Agency; neither the European Union nor the granting authority can be held responsible for them. NC acknowledges the financial support from the Royal Society (grant number URFR1251326).

\begin{appendix}
\numberwithin{equation}{section}
\section{JNPT transition to the saturated ferromagnet}
\label{app:PT}
In order to describe the JNPT transition it is convenient to change variables
\be
\psi=1-2\rho\ ,
\ee
and then define a complex Bose field by
\be
\Phi=\sqrt{2S\rho}e^{-i\theta}\ .
\ee
We want to think of approaching the transition by tuning $\lambda_2$ to its critical value $\lambda_{2,c}$ at fixed $\lambda_4$. Close to the transition the magnon density $\langle\rho\rangle$ then tends to zero. Substituting the above field redefinitions into \fr{SFT} we obtain
\be
\mathcal{L}=\Phi^\dagger(\partial_\tau-JS\partial_x^2-\mu)\Phi+c(\Phi^\dagger)^2\Phi^2+\dots\ .
\label{LLL}
\ee
Here $c >0$ and the chemical potential shows the characteristic square root dependence of the deviation from the JNPT transition
\be
\mu\propto \sqrt{\lambda_2-\lambda_{2,c}}\ .
\ee
Eqn \fr{LLL} is the imaginary-time Lagrangian density of the Lieb-Liniger model \cite{LiebLiniger1963}. We therefore can use the exact results for the latter to determine the scaling of the Luttinger parameter close to the transition
\be
K-1\propto\sqrt{\mu}\ \ ,\quad \mu\to 0.
\label{KJNPT}
\ee
\end{appendix}

\bibliography{bibliography.bib}


\end{document}